\documentstyle[aps,prl,multicol,epsfig,amssymb,amsmath]{revtex}

\title{Elastic Interactions of Cells}
\author{U. S. Schwarz$^{1}$ and S. A. Safran$^{2}$}
\address{$^{1}$Max-Planck-Institute of Colloids and Interfaces,
14424 Potsdam, Germany \\
$^{2}$Department of Materials and Interfaces,
The Weizmann Institute of Science, Rehovot 76100, Israel}   

\def\eq#1{Eq.~(\ref{#1})}
\def\fig#1{Fig.~\ref{#1}}

\def\kd{\delta_{ij}}

\def\r{\textbf{r}}
\def\s{\textbf{s}}

\def\P{\textbf{P}}
\def\f{\textbf{f}}

\def\u{\textbf{u}}
\def\n{\textbf{n}}

\begin{document}

\maketitle

\begin{abstract}
  Biological cells in soft materials can be modeled as anisotropic
  force contraction dipoles.  The corresponding elastic interaction
  potentials are long-ranged ($\sim 1/r^3$ with distance $r$) and
  depend sensitively on elastic constants, geometry and cellular
  orientations.  On elastic substrates, the elastic interaction is
  similar to that of electric quadrupoles in two dimensions and for
  dense systems leads to aggregation with herringbone order on a
  cellular scale.  Free and clamped surfaces of samples of finite size
  introduce attractive and repulsive corrections, respectively, which
  vary on the macroscopic scale.  Our theory predicts cell
  reorientation on stretched elastic substrates.
\end{abstract}

% PACS numbers:
% 87.10.+e General theory and mathematical aspects of biological physics 
% 61.72.Ji Point defects and defect clusters 
% 87.18.Hf Spatiotemporal pattern formation in cellular populations

\begin{multicols}{2}
  
  Biological cells can exert strong physical forces on their
  surroundings. One example are fibro\-blasts, which are mechanically
  active cells found in connective tissue.  In the early 1980s, Harris
  and coworkers found that fibroblasts exert much more force than
  needed for locomotion \cite{c:harr80}.  They suggested that strong
  fibroblast traction is needed in order to align the collagen fibers
  in the connective tissue. Since cell locomotion is guided by
  collagen fibers, this results in a \emph{mechanical interaction} of
  cells.  The interplay of fiber alignment and cell locomotion has
  been analyzed theoretically in the framework of coupled transport
  equations for fiber and cell degrees of freedom \cite{c:oste83}.
  However, it is well known that cellular behavior is also affected by
  purely elastic effects, which were not considered in these studies.
  For example, stationary cells plated on an elastic substrate which
  is cyclically stretched reorientate away from the stretching
  direction \cite{c:dart86}, and locomoting cells on a strained
  elastic substrate reorientate in the strain direction \cite{c:lo00}.
  Recent experiments show that adhering cells sense mechanical signals
  through focal adhesions \cite{c:bala01}.  In contrast to chemical
  diffusion fields, elastic effects are long-ranged and propagate
  quickly, and they are known to be important during development,
  wound healing, inflammation and metastasis \cite{c:chic98}.
  
  In this Letter, we consider theoretically the possibility of
  \emph{elastic interaction} of cells. We focus on static forces, a
  situation which should apply to cells with restricted cytoskeletal
  regulation or to artificial cells which have a biomimetic
  contractile system without any regulation; the theoretical framework
  presented here for this case is a prerequisite for understanding the
  more complicated cases, e.g.\ the case of locomoting cells with a
  regulated response and dynamic force patterns \cite{c:lo00}.  In the
  static case, the elastic interaction of cells through their strain
  fields leads to forces and torques which can change their positions
  and orientations. If the cellular configuration can relax to
  equilibrium, the final configuration will be a minimum of the
  elastic energy. In the following, we derive the laws for elastic
  interactions of cells (which are modeled as anisotropic force
  contraction dipoles) and show how they depend on elastic constants,
  distance, cellular orientations, geometry and boundary conditions.

  If the distance between cells is much larger than their spatial
  extent, they can be modeled as point defects in an elastic medium.
  Elastic interactions of point defects have been discussed before for
  e.g.\ hydrogen in metal \cite{e:wagn74}, atoms adsorbed onto crystal
  surfaces \cite{e:lau77}, and graphite intercalation compounds
  \cite{e:safr79}. For each defect, the force is restricted to a small
  region of space, and the force distribution can be characterized by
  its \emph{force multipoles} \cite{e:siem68}
\begin{equation} \label{eq:force_multipoles}
  P_{i_1 \dots i_n i} = \int s_{i_1} \dots s_{i_n}\ f_i(\s)\ d\s
\end{equation}
where $\f$ is the force density.  The force monopole $\P$ is the
overall force, which vanishes for inert particles. Therefore in the
classical case, the first relevant term is the force dipole $P_{ij}$,
which describes the dilating/contracting action of the force
distribution and has the dimension of an energy. Previous studies of
elastic interactions of force multipoles were mostly concerned with
\emph{isotropic} force \emph{dilation} dipoles (that is $P_{ij} = P
\kd$ with $P > 0$) and the finite sample size effect of \emph{free}
surfaces.  The biological case which we discuss here is different in
several respects.  First, since cells can act as active walkers, there
exists the possibility of force monopoles. Second, cellular force is
based on actomyosin contractility and therefore leads to force
\emph{contraction} dipoles (that is $P < 0$). Third, adhering cells in
most cases generate highly \emph{anisotropic} force patterns, that is,
the force dipole is not isotropic and will reorientate with respect to
the surrounding strain field (e.g.\ on strained substrates).  And
fourth, in biological cases the elastic medium (e.g.\ the tissue) has
\emph{clamped} rather than free surfaces. In fact it is well known
that cells become mechanically active only if their environment can
support enough stress, thus clamped boundary conditions are often
needed to induce cellular activiation \cite{c:tran99}.

We assume that the elastic medium of interest (real or artificial
tissues, elastic substrates) propagates strain like an isotropic
elastic medium with a Young modulus in the order of kPa. For elastic
substrates, the Poisson ratio $\nu$ is close to $1/2$ (incompressible
case). For the following it is convenient to define $\Lambda = \lambda
/ \mu$ and $c = 2 \mu + \lambda = \mu (2+\Lambda)$, where $\mu$ and
$\lambda$ are the Lam\'{e} coefficients of the isotropic elastic
medium. The incompressible case $\nu = \Lambda / 2 (\Lambda + 1) =
1/2$ then corresponds to the limit $\Lambda \to
\infty$ with $\Lambda / c \to 1 / \mu$. Propagation of strain in an
infinite isotropic elastic medium is described by the Green function
\cite{b:land70}
\begin{equation} \label{eq:Green_3d}
G^{3d}_{ij} = \frac{1}{8 \pi c} \left\{ 
(3+\Lambda) \kd + (1+\Lambda) \frac{x_i x_j}{r^2} \right\} \frac{1}{r}
\end{equation}
where $r$ denotes the distance from the force center.  For cells
plated on an elastic substrate with a substrate thickness that is much
larger than the displacements caused by cell traction, the relevant
Green function is the one of an isotropic elastic halfspace with a
free surface. Since such cells apply only tangential traction, we need to
specify $G_{ij}$ only for the x-y-plane \cite{b:land70}:
\begin{equation} \label{eq:Green_2d}
G^{2d}_{ij} = \frac{(2 + \Lambda)}{4 \pi (1+\Lambda) c} \left\{ 
(2+\Lambda) \kd + \Lambda \frac{x_i x_j}{r^2} \right\} \frac{1}{r}\ .
\end{equation}
The elastic interaction energy $W$ between two force distributions can
be written as a function of their force multipoles and the Green
function \cite{e:siem68}
\begin{align}
W & = - \int \int f_i(\s)\ G_{ij}(|\s - \s'|)\ f_j(\s')\ d\s\ d\s' \nonumber \\
\label{eq:siems}
  & = - \sum_{n = 0}^{\infty} \sum_{m = 0}^{\infty} \frac{1}{n!} \frac{(-1)^m}{m!} 
G_{ij,i_1 \dots i_n j_1 \dots j_m} P_{i_1 \dots i_n i} P'_{j_1 \dots j_m j} 
\end{align}
where we sum over repeated indices and where indices after the comma
represent partial derivatives. In the second line, the first line has
been expanded twice and the definitions of \eq{eq:force_multipoles}
have been used.

The interaction between two force monopoles $\P$ and $\P'$ at
$\r$ and $\r'$, respectively, follows from \eq{eq:siems} as $W = - P_i
G_{ij}(\r-\r') P'_j$. In the incompressible limit, this can be written
as $W = - (\P \cdot \P' + (\P \cdot \n) (\P'
\cdot \n)) / 8 \pi \mu r$, where $\n$ is the normalized separation
vector between the two monopole locations (in three dimensions; in two
dimensions, an additional factor of $2$ appears). This interaction is
similar to the one between electric dipoles \cite{tlus00}, thus we
expect chaining to dominate large scale assembly, as confirmed by
Monte Carlo simulations (not shown). However, since the force
monopoles correspond to active movements, a model for cell locomotion
is required to fully treat this case.

It is generally accepted that mechanically active cells exert only a
very small overall force. Moreover, in most cases they are usually
found to have highly polarized, that is, pinching force patterns
\cite{c:demb99}. In the following we therefore 
model cellular force patterns as anisotropic force contraction
dipoles. The direction $\hat{\r}$ of the pinch can be extracted from
experimentally measured force patterns by determining the direction
of the eigenvector of the force dipole tensor corresponding to its
largest eigenvalue. Then the force dipole tensor can be approximated
as $P_{ij} = P \hat{r}_i \hat{r}_j$.  In many cases, the cell
orientation following from the force pattern corresponds to the cell
orientation following from overall cell shape or staining for actin
fibers.  For both locomoting and stationary fibroblasts on elastic
substrates, the magnitude of the force dipole can be estimated to be
of the order of $P \approx - 10^{-11} J$ (this corresponds to a
pinching pair of forces, separated by a distance of 60 $\mu$m and each
200 nN strong). The corresponding length scale (e.g.\ for
displacements close to the cell) is $(P / c)^{1/3} \approx 10\ \mu$m,
which is somewhat smaller than a typical cell size ($\approx 50\
\mu$m).  The interaction between two force dipoles $P_{li}$ and
$P'_{kj}$ at $\r$ and $\r'$, respectively, follows from
\eq{eq:siems} as
\begin{equation} \label{eq:interaction}
W(\r,\r') = - P_{li} u_{i,l}(\r,\r') = P_{li} G_{ij,lk}(\r-\r') P'_{kj}
\end{equation}
where $\u(\r,\r')$ is the displacement at $\r$ produced by the force
dipole at $\r'$. Since $G \sim 1/r$, the elastic interaction between
force dipoles scales as $\sim 1/r^3$. 

If the cells have isotropic force dipoles, their elastic interactions
are well known: in infinite space, $W = P^2 G^{3d}_{ij,ij} = 0$
\cite{e:siem68} and an elastic interaction can only by induced by the
boundary conditions \cite{e:wagn74}. On a semi-infinite space with
free surface, $W = P^2 G^{2d}_{ij,ij} = (2+\Lambda)^2 P^2 / 4 \pi
(1+\Lambda) c r^3$, thus the interaction is isotropic and repulsive
\cite{e:lau77}. However, in most cases the cells will have highly
anisotropic force dipoles.  We start with the half\-space and consider
the following situation: one of the two interacting dipoles is fixed
at the origin with vanishing polar angle. The other dipole is a
distance $r$ away with polar angle $\alpha$.  The polar angle of the
separation vector is denoted by $\beta$. Using \eq{eq:Green_2d} in
\eq{eq:interaction}, we find
\begin{equation} \label{eq:interaction_2d}
W(r,\alpha,\beta) = \frac{(2+\Lambda) P^2}{4 \pi c (1+\Lambda) r^3} f(\alpha,\beta)
\end{equation}
with
\begin{align} 
f(\alpha,\beta) = & \frac{1}{8} [ (4 + 3 \Lambda) \cos (2 \alpha) 
  + 15 \Lambda \cos (2 ( \alpha - 2 \beta )) \nonumber \\
\label{eq:interaction_2d_2}
+ (2 + \Lambda) & (2 + 6 \cos (2 ( \alpha - \beta ) ) + 6 \cos (2 \beta)) ]\ .
\end{align}
Depending on orientation, the interaction can be repulsive or
attractive.  The attractive component leads to orientation dependent
aggregation.

\begin{figure}[h]
\begin{center}
\epsfig{figure=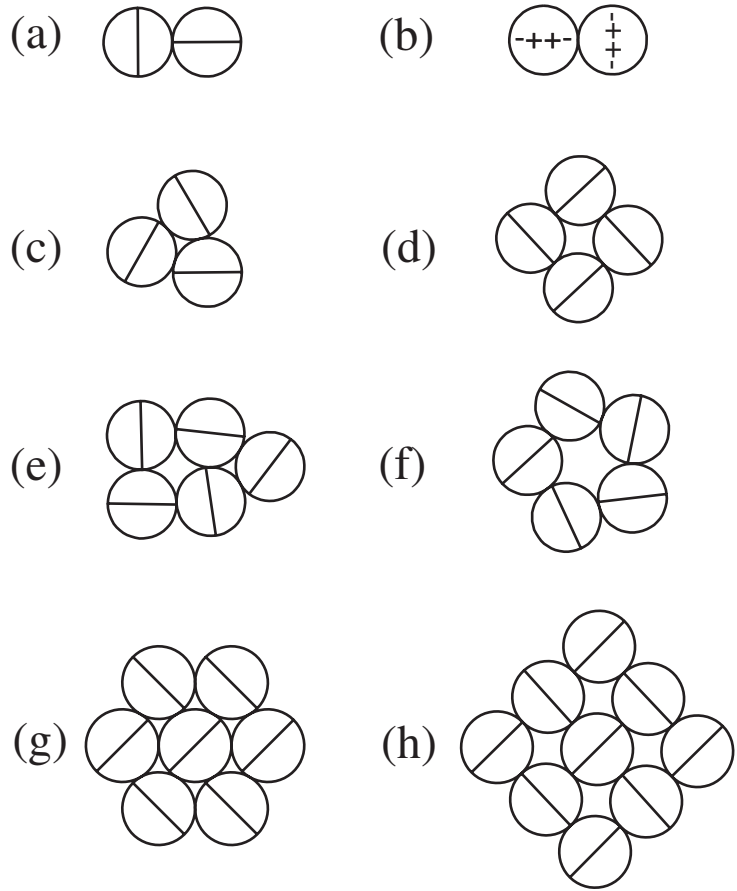}
\end{center}
\caption{Typical configurations for anisotropic force dipoles on an 
  incompressible elastic substrate. (a) Minimal energy configuration
  for two dipoles (f = -2).  (b) Linear electric quadrupoles have the
  same optimal configuration.  (c) - (f) Small clusters are subject to
  surface reconstruction. For five particles, (e) and (f) are nearly
  degenerate (f = -11.59 and f = -11.43, respectively).  (g) At high
  densities, herringbone order results. (h) In terms of energy, the
  square lattice is most favorable.}
\label{fig:SmallClusters}
\end{figure}

In order to investigate this point in more detail, we consider force
dipoles with a spherical hard core (corresponding to a typical cell
size). For $\Lambda \approx 0$ (vanishing Poisson ratio), the only
favorable alignments will be side-by-side and the cells will assemble
into linear strings, with their orientations perpendicular to the
string direction. For larger $\Lambda$, aggregation will be much more
compact.  In the incompressible case, for a given angle $\beta$ the
optimal angle $\alpha$ follows as $\alpha_{min} \approx \pi / 2 + 2
\beta$; the corresponding $f$ varies between -2 and -1.4. The optimal
energy -2 is obtained for the four perpendicular
configurations. Considering only nearest neighbor interactions would
lead to a square lattice at higher densities. However, at area
densities beyond $\pi / 4$, this structure is overpacked and a
herringbone structure results. Although the herringbone structure does
not achieve the lower energy values of the square lattice, it can
exist up to area density $\pi / 2 \sqrt{3}$ and will be favored for
entropic reasons. For finite-sized clusters, surface
reconstruction will take place. Moreover, for an increasing number of
particles the interaction of \eq{eq:interaction_2d} leads to an
increasingly rugged energy landscape with many local minimia due to
the long range and orientation dependance of the potential. As a
result of this metastability, unusual patterns like rings will form
for certain initial conditions. In \fig{fig:SmallClusters} we show
some typical configurations.

It is instructive to note that the orientational part of the
interaction in the incompressible limit is very similar to that of
linear electric quadrupoles in two dimensions (see
\fig{fig:SmallClusters}).  This analogy is due to the fact that
the corresponding interaction energy for linear electric quadrupoles,
$W = G^{el}_{,ijkl} P_{ij} P'_{kl}$ (where $G^{el}(r) \sim 1/r$ is the
electric Green function), arises from contracting tensors of analogous
symmetry. However, the near perfect agreement in the angle-dependent
part is an accidental result for the incompressible case; moreover,
electric quadrupoles interact with $1/r^5$ rather than with
$1/r^3$.

We now discuss the elastic interaction of anisotropic force contraction
dipoles acting in a three-dimensional elastic medium, which follows by
using \eq{eq:Green_3d} in \eq{eq:interaction}. This interaction is
more complicated than the two-dimensional one, since its
orientation dependance involves three rather than two different
angles. A detailed discussion will be given elsewhere. Here we only
discuss some high symmetry cases: for two parallel dipoles pointing in
z-direction and placed along the x-axis, we find
\begin{equation} \label{eq:interaction_3d}
W^{direct}(x) = \frac{(\Lambda - 1) P^2}{8 \pi c x^3}\ .
\end{equation}
Thus this interaction changes sign as $\Lambda$ varies through $1$
($\nu = 1/4$). If one considers force dipoles
arranged around a central dipole in the x-z-plane, similar
considerations apply as in the two-dimensional case: for example, when
considering only nearest neighbor interactions, a square arrangement
of perpendicular dipoles with interaction energy $W(r) = - (\Lambda +
1) P^2 / 4 \pi c r^3$ is the most favorable one. However, if one now
tries to continue this arrangement in the third dimension, frustration
effects result. Depending on initial conditions, this enhances the
occurance of irregular patterns.

\begin{figure}[h]
\begin{center}
\epsfig{figure=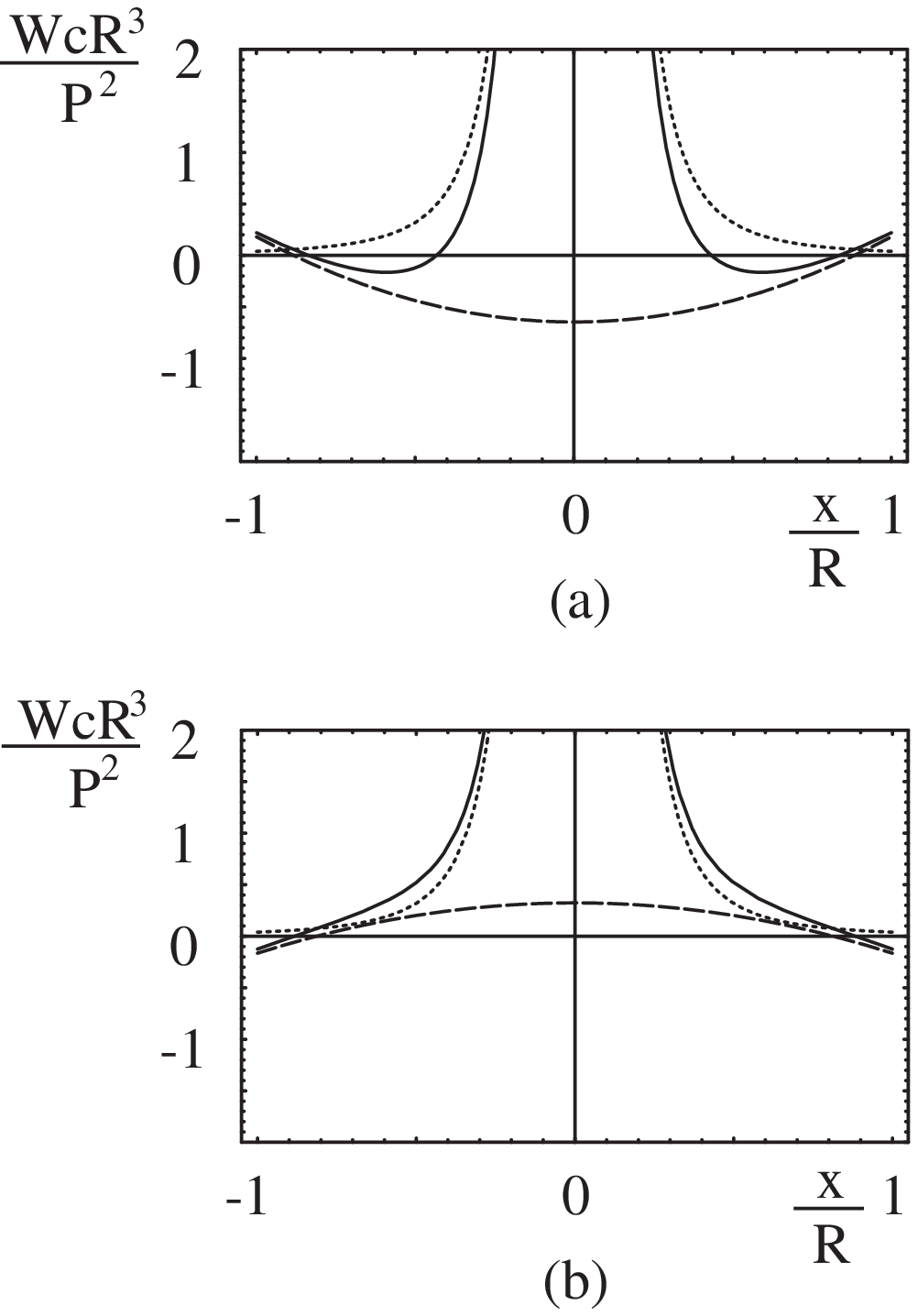}
\end{center}
\caption{Elastic interaction energy $W$ in units of $P^2 / c R^3$ for 
  two parallel anisotropic force dipoles of magnitude $P$ in an
  isotropic elastic sphere with radius $R$ and elastic constants $c$
  and $\Lambda = 2$ (Poisson ratio $\nu = 1/3$). For this value of
  $\Lambda$, the direct elastic interaction is repulsive (dotted
  lines). (a) Free surface: the image correction (dashed line) is
  attractive and generates a new minimum in the full interaction
  potential (solid line). (b) Clamped surface: the image correction is
  repulsive.}
\label{fig:ElasticSphere}
\end{figure}

Due to the long-ranged nature of the elastic interaction, boundary
effects will be very important \cite{e:wagn74,e:zabe79}. As an
instructive example, we now discuss the elastic interaction of
anisotropic force contraction dipoles in an isotropic elastic sphere
of macroscopic radius $R$. For a free surface, this situation has been
investigated before for both isotropic \cite{e:wagn74} and anisotropic
force dipoles \cite{e:hirs81}. However, no such treatment exists for
clamped surfaces, which are expected to have larger biological
relevance. For both free and clamped surfaces, one has to introduce
image displacements, which can be determined using expansions in vector
spherical harmonics \cite{e:hirs81}.  Again the general expressions
will be given elsewhere, and here we only consider the high symmetry
case of two dipoles both orientated in z-direction. Their direct
interaction along the x-axis is given by \eq{eq:interaction_3d}. The
image interaction follows from inserting the image displacements of
the first dipole (which are complicated functions expressed as vector
spherical harmonics) into \eq{eq:interaction}. For simplicity, here we
report only the results for the imcompressible limit.  We find
\begin{equation}
\label{eq:freesurface}
W^{img}_{free}(x) = \frac{P^2}{76 \pi \mu R^3} (-45 + 48 \left( \frac{x}{R} \right)^2)
\end{equation}
for a free surface and
\begin{equation}
\label{eq:clampedsurface}
W^{img}_{clamped}(x) = \frac{P^2}{20 \pi \mu R^3} (2 - 15 \left( \frac{x}{R} \right)^2)
\end{equation}
for a clamped surface. Therefore free and clamped surfaces introduce
attractive and repulsive corrections, respectively.  In
\fig{fig:ElasticSphere} we show the interaction energies $W$ for
$\Lambda = 2$ ($\nu = 1/3$) as a function of distance $x$. We see that
the image corrections can produce new minima in the full interaction
potential.  The main conclusion from
Eqs.~(\ref{eq:freesurface},\ref{eq:clampedsurface}) is the fact that
the image effects lead to corrections which operate on the macroscopic
scale $R$. For the case of hydrogen in metal, this is known to lead to
structure formation on a macroscopic scale (\emph{macroscopic modes})
\cite{e:wagn74,e:zabe79}. In the biological case, the boundary induced
pattern formation competes with structure formation on cellular and
elastic scales, which results from the direct elastic interaction.
Therefore we expect that in a theory for cellular densities modes in
elastic media of finite size, hierarchical structures will result.
Like for hydrogen in metal, image effects can be also expected to lead
to incoherent deformations (fracture).

The simplest case of the interaction of a cellular force pattern with
an elastic strain field is the case of a single cell plated on an
elastic substrate which is homogeneously stretched along the
x-direction by applying stress $p > 0$ at the sides. Then $(u_x, u_y)
= (p/E) (x, - \nu y)$, with $E$ being the Young modulus. The
interaction energy follows as $W = - P_{li} u_{i,l} = - (P p / E)
(\cos^2 \alpha - \nu \sin^2 \alpha)$, where $\alpha$ is the polar
angle describing cell orientation. Since $P < 0$, the cell will
reorientate \emph{perpendicular} to the direction of stretching
($\alpha = 90^\circ$). In the case of compression, $p < 0$, the cell
will reorientate \emph{parallel} to the direction of compression
($\alpha = 0^\circ$).  Although the dynamic aspects of elastic
interactions of cells are out of the scope of the present work, it is
worth noting that on cyclically stretched elastic substrates, $p$
periodically changes sign and in order to maintain a stationary state,
the cell might try to avoid both tensile and compressive strain.  One
easily calculates that this corresponds to $\alpha = \arccos \sqrt{\nu
/ (1 + \nu)}$.  For $\nu \approx 0.4$, this yields $\alpha \approx
60^\circ$, in excellent agreement with experiments \cite{c:dart86}.

\textbf{Acknowledments:} It is a pleasure to thank N. Q. Balaban, A.
Bershadsky, P. Fratzl, B. Geiger, D. Riveline and T. Tlusty for helpful
discussions. USS thanks the Minerva Foundation and the Emmy Noether
Program of the German Science Foundation for support. SAS thanks
the Schmidt Minerva Center and the Center on Self-Assembly sponsored
by the Israel Science Foundation.

%\bibliographystyle{unsrt}
%\bibliography{a,b,c,e,uss}  

\end{multicols}

\newpage

\newpage          

\end{document}